 \documentclass[12pt,preprint]{aastex}

 \usepackage{epsfig}

\newcommand{\rs}{r_{_{\rm S}}}

\newcommand{\enrate}{\dot E}
\newcommand{\momrate}{\dot J}

\newcommand{\ellK}{\ell_{\rm K}}

\newcommand{\gapprox}{\lower.4ex\hbox{$\;\buildrel >\over{\scriptstyle\sim}\;$}}
\newcommand{\lapprox}{\lower.4ex\hbox{$\;\buildrel <\over{\scriptstyle\sim}\;$}}
\newcommand{\begeq}{\begin{equation}}
\newcommand{\fineq}{\end{equation}}
\newcommand{\begeqarray}{\begin{eqnarray}}
\newcommand{\fineqarray}{\end{eqnarray}}
\newcommand{\green}{f_{_{\rm G}}}
\newcommand{\Qjet}{L_{\rm jet}}
\newcommand{\Qesc}{L_{\rm esc}}

\newcommand{\Nesc}{\dot N_{\rm esc}}

\def\ellprime0{\ell'_0}

\slugcomment{accepted by ApJ Letters}

\shorttitle{Shocks in Viscous ADAF Disks}

\shortauthors{Becker, Das, \& Le}

\begin{document}

\title{Particle Acceleration and the Formation of
Relativistic Outflows in Viscous Accretion Disks with Shocks}

\author{Peter A. Becker,\altaffilmark{1} Santabrata Das,\altaffilmark{2}
and Truong Le\altaffilmark{3}}

\altaffiltext{1}{College of Science, George Mason University, Fairfax, VA
22030-4444; pbecker@gmu.edu}
\altaffiltext{2}{Astrophysical Research Center, Sejong University, Seoul 143-747,
Korea; sbdas@canopus.cnu.ac.kr}
\altaffiltext{3}{E. O. Hulburt Center for Space Research, Naval Research
Laboratory, Washington, DC 20375; tle@ssd5.nrl.navy.mil}

\begin{abstract}
In this Letter, we present a new self-consistent theory for the
production of the relativistic outflows observed from radio-loud black
hole candidates and active galaxies as a result of particle acceleration
in hot, viscous accretion disks containing standing,
centrifugally-supported isothermal shocks. This is the first work to
obtain the structure of such disks for a relatively large value of the
Shakura-Sunyaev viscosity parameter ($\alpha=0.1$), and to consider the
implications of the shock for the acceleration of relativistic particles
in viscous disks. In our approach, the hydrodynamics and the particle
acceleration are coupled and the solutions are obtained
self-consistently based on a rigorous mathematical method. We find that
particle acceleration in the vicinity of the shock can provide enough
energy to power the observed relativistic jet in M87.

\end{abstract}



\section{INTRODUCTION}

It has recently been established that the acceleration of relativistic
particles at a standing shock in an advection-dominated accretion flow
(ADAF) can power the outflows frequently observed from radio-loud active
galactic nuclei (AGNs) and galactic black-hole candidates (Le \& Becker
2004, 2005, 2007). Radio-loud AGNs are thought to contain supermassive
central black holes surrounded by hot, two-temperature ADAFs with
significantly sub-Eddington accretion rates. In these disks, the ion
temperature $T_i \sim 10^{12}\,$K greatly exceeds the electron
temperature $T_e \sim 10^{10}\,$K (e.g., Narayan, Kato, \& Honma 1997;
Becker \& Le 2003). The observed correlation between high radio
luminosities and the presence of the outflows suggests that hot ADAF
disks are able to efficiently accelerate the relativistic particles
powering the jets. In fact, such disks are ideal sites for first-order
Fermi acceleration at shocks because the gas is tenuous, and therefore a
significant fraction of the accelerated particles are able to avoid
thermalization and escape from the disk. Although the work of Le \&
Becker (2004, 2005, 2007) was the first to establish a direct connection
between the structure of the disk/jet system and a specific
microphysical particle acceleration mechanism, their model was only
applicable to fully inviscid (adiabatic) disks. In this Letter, we
generalize their model to include viscous dissipation.

\section{DYNAMICAL MODEL}

The dynamical structure of the accretion disks considered here is based
on the viscous inflow model studied by Narayan et. al (1997), Lu, Gu, \&
Yuan (1999), Gu \& Lu (2001), and Becker \& Le (2003), with the addition
of an isothermal shock. In the one-dimensional, vertically-integrated,
steady state ADAF scenario under consideration here, the accretion rate
$\dot M$ and the angular momentum transport rate $\momrate$ are
conserved, where
\begeq
\dot M \equiv 4 \pi r H \rho \, u
\ , \ \ \ \
\momrate \equiv \dot M \, \ell - {\cal G}
\ ,
\label{eq1}
\fineq
with $\rho$ denoting the mass density, $u$ the radial velocity (defined
to be positive for inflow), $H$ the disk half-thickness, $\ell=r^2 \,
\Omega$ the specific angular momentum, $\Omega$ the angular velocity,
and ${\cal G}$ the torque. The vertical hydrostatic structure of the
disk is described by the usual relations
\begeq
H(r) = {r^2 a\over \ellK}
\ , \ \ \
a^2(r) = {P \over \rho}
\ ,
\label{eq2}
\fineq
where $a$ denotes the isothermal sound speed, and $\ellK$ represents the
Keplerian angular momentum per unit mass for matter orbiting in the
pseudo-Newtonian potential $\Phi$, given by (Paczy\'nski \& Wiita 1980)
\begeq
\ellK^2(r) \equiv {GM \, r^3 \over (r-\rs)^2}
= r^3 \, {d\Phi \over dr}
\ ,
\label{eq3}
\fineq
with $\rs \equiv 2\,GM/c^2$ denoting the Schwarzschild radius for a
black hole of mass $M$. The energy transport rate
\begeq
\enrate = - {\cal G}\, {\ell \over r^2} + \dot M \left(
{1 \over 2}\,{\ell^2 \over r^2} + {1\over 2} \, u^2
+ {P + U \over \rho} + \Phi\right)
\ ,
\label{eq4}
\fineq
is also conserved (except at the shock, if one is present), where $U$ is
the internal energy density, $P$ is the gas pressure, and all quantities
represent vertical averages. We assume that the ratio of specifics
heats, $\gamma \equiv (U + P) / U$, remains constant throughout the
flow. Note that the transport rates $\dot M$, $\momrate$, and $\enrate$
are all defined to be positive for inflow.

In a steady state, the comoving radial acceleration rate in the frame of
the accreting gas is expressed by
\begeq
{Du \over Dt} \equiv
- u \, {du \over dr} = {1\over \rho} {dP \over dr}
+ {\ellK^2 - \ell^2 \over r^3}
\ ,
\label{eq5}
\fineq
and the torque ${\cal G}$ is related to the gradient of the angular
specific momentum $\ell$ via (e.g., Frank, King, \& Raine 1985)
\begeq
{\cal G} = - 4 \pi r H \rho \, \nu \,
\left({d\ell \over dr} - {2 \ell \over r}\right) \ ,
\ \ \ \ 
\nu = {\alpha \, r^2 a^2 \over \ellK}
\ ,
\label{eq6}
\fineq
where $\nu$ is the kinematic viscosity, computed using the standard
Shakura-Sunyaev (1973) prescription, with constant $\alpha$. Away from
the shock location, the variation of the internal energy density is
governed by viscous dissipation and adiabatic compression, and the
comoving rate of change of $U$ is therefore given by (e.g., Becker \& Le
2003)
\begeq
{DU \over Dt} \equiv
- u \, {dU \over dr} = - \gamma \, {U \over \rho} \, u \,
{d\rho \over dr} + {\rho \nu \over r^2}
\left({d\ell \over dr} - {2 \ell \over r}\right)^2
\ .
\label{eq7}
\fineq
By combining various relations, one can obtain the differential
dynamical equation (e.g., Narayan et. al 1997)
\begeq
\left({u^2 \over a^2} - {2 \, \gamma \over \gamma+1}\right)
{d\ln u \over dr} = {\ell^2 - \ellK^2 \over a^2 \, r^3}
+ {2 \, \gamma \over \gamma+1} \left({3 \over r} - {d\ln\ellK
\over dr}\right) + \left(\gamma-1 \over \gamma+1\right) {u \,
\ellK \, (\ell - j)^2 \over \alpha \, a^4 r^4}
\ ,
\label{eq8}
\fineq
where $j\equiv \dot J/\dot M$. This expression is supplemented by the
differential conservation equation for the specific angular momentum,
\begeq
{d \ell \over dr} = {2 \, \ell \over r} - {u \, \ellK \,
(\ell - j) \over \alpha \, r^2 a^2}
\ ,
\label{eq9}
\fineq
obtained by utilizing equations~(\ref{eq1}) and (\ref{eq6}). Dynamical
solutions are computed by simultaneously integrating
equations~(\ref{eq8}) and (\ref{eq9}). In order to ensure the stability
of the calculations, the integrations are performed in the outward
direction, starting with initial values near the event horizon computed
using the boundary conditions derived by Becker \& Le (2003).

The resulting disk/shock model depends on several parameters, namely the
energy transport rate per unit mass, $\epsilon\equiv \dot E/\dot M$, the
angular momentum transport rate, $j\equiv \dot J/\dot M$, the ratio of
specific heats, $\gamma$, and the viscosity parameter, $\alpha$. The
value of $\epsilon$ is constant in ADAF disks except at the shock
location. When a shock is present, we use the subscripts ``-'' and ``+''
to refer to quantities measured just upstream and just downstream from
the shock, respectively. Critical points occur where the left- and
right-hand sides of equation~(\ref{eq8}) vanish simultaneously. The flow
must pass through at least one critical point before crossing the event
horizon since general relativity requires supersonic inflow at the
horizon. If the flow is smooth (shock-free), then the gas passes through
only one critical point, located at radius $r=r_c$. If a shock is
present in the flow, then the gas passes through one critical point in
the pre-shock region at $r=r_c^{\rm out}$, and through another in the
post-shock region at $r=r_c^{\rm in}$ (Abramowicz \& Chakrabarti 1990).

The isothermal shock radius, $r_*$, must be determined self-consistently
by satisfying the velocity and energy jump conditions (Chakrabarti 1989)
\begeq
{u_+ \over u_-} = {1 \over {\cal M}_-^2}
\ , \ \ \ \ \
\Delta\epsilon \equiv \epsilon_+-\epsilon_- = {u_+^2 - u_-^2 \over 2}
\ ,
\label{eq10}
\fineq
where ${\cal M}_- \equiv u_-/a_-$ is the upstream Mach number at the
shock location. Note that the velocity jump condition we employ is
slightly different from the one utilized by Le \& Becker (2005) because
those authors defined the Mach number in terms of the adiabatic sound
speed rather than the isothermal sound speed used here. Due to the
escape of energy at the isothermal shock location, the energy transport
rate $\epsilon$ drops from the upstream value $\epsilon_-$ to the
downstream value $\epsilon_+$, and consequently $\Delta \epsilon < 0$.
The power lost from the disk at the isothermal shock is related to the
observed jet kinetic luminosity, $\Qjet$, via
\begeq
\Qjet = - \dot M \, \triangle \epsilon
\ ,
\label{eq11}
\fineq
where $\dot M$ is the accretion rate computed using the observed total
energy output for a specific source, and the negative sign appears because
$\Delta \epsilon < 0$.

For given observational values of $M$, $\dot M$, and $\Qjet$, the
process of determining the structure for a disk containing an isothermal
shock begins with the selection of provisional values for the energy
inflow rate $\epsilon_-$ and the angular momentum inflow rate $j$. Next
we numerically integrate equations~(\ref{eq8}) and (\ref{eq9}), starting
from a location near the horizon and working outward towards the inner
critical point. Once a solution is established that passes smoothly
through the inner critical point, the next step is to determine the
shock location $r_*$ by ensuring that the shock jump conditions
(eqs.~[\ref{eq10}]) are satisfied. However, if the resulting value of
$\Delta\epsilon$ is not consistent with equation~(\ref{eq11}), then the
values of $\epsilon_-$ and $j$ are adjusted and the procedure is
repeated starting with the integration from the horizon. When this step
is successfully completed, the integration is continued from the
upstream side of the shock outward toward the outer critical point. If
the flow does not pass smoothly through the outer critical point, then
the values of $\epsilon_-$ and $j$ are modified and the procedure is
repeated starting from the horizon. The end result is a unique set of
values for $\epsilon_-$ and $j$, along with the associated global
solution for the disk/shock structure.

\section{PARTICLE TRANSPORT EQUATION}

In a steady-state situation, the Green's function, $\green(E,r)$,
representing the particle distribution resulting from the continual
injection of monoenergetic seed particles with energy $E_0$ from a
source located at the shock radius $r_*$ satisfies the
vertically-integrated transport equation (Le \& Becker 2005, 2007)
\begeqarray
- H u {\partial \green \over \partial r} &=& - {1 \over 3 r}
\, {d \over dr} (r H u) \, E \, {\partial \green
\over \partial E} + {1 \over r} {\partial \over \partial r}
\left(r H \kappa \, {\partial \green \over \partial r}\right)
\nonumber \\
&+& {\dot N_0 \, \delta(E-E_0) \, \delta(r-r_*) \over (4 \pi E_0)^2 r_*}
- A_0 \, c \, H_* \, \delta(r-r_*) \, \green
\ ,
\label{eq12}
\fineqarray
where $H_*$ denotes the disk half-thickness at the shock location,
$\kappa$ is the spatial diffusion coefficient, $\dot N_0$ denotes the
particle injection rate, $c$ is the speed of light, and $A_0$ is a
parameter that describes the rate of particle escape through the upper
and lower surfaces of the disk. The left-hand side of
equation~(\ref{eq12}) represents the co-moving (advective) time
derivative and the terms on the right-hand side describe first-order
Fermi acceleration, spatial diffusion, the particle source, and the
escape of particles from the disk, respectively. The relativistic
particle number and energy densities associated with the Green's
function are given by $n(r)=\int_0^\infty 4 \pi E^2 \, \green \, dE$ and
$U(r)=\int_0^\infty 4 \pi E^3 \, \green \, dE$, respectively. Although
the Fermi acceleration of the particles is concentrated at the shock,
the rest of the disk also contributes to the particle acceleration
because of the general convergence of the MHD waves in the accretion
flow.

After the velocity profile has been determined, we can compute the
Green's function describing the energy and space distribution of the
accelerated relativistic particles inside the disk by solving
equation~(\ref{eq12}). We set the injection energy using $E_0 =
0.002\,$ergs, which corresponds to an injected Lorentz factor $\Gamma_0
\equiv E_0 / (m_p \, c^2) \sim 1.3$, where $m_p$ is the proton mass. The
speed of the injected particles, $v_0 = c \, (1-\Gamma_0^{-2})^{1/2}$,
is about three to four times higher than the mean ion thermal velocity
at the shock location. The seed particles are picked up from the
high-energy tail of the Maxwellian distribution for the thermal ions.
The particle injection rate $\dot N_0$ is computed using the energy
conservation condition (cf. eq.~[\ref{eq11}])
\begeq
\dot N_0 \, E_0 = - \dot M \, \Delta \epsilon = \Qjet
\ .
\label{eq13}
\fineq
This self-consistency relation ensures that the energy injection rate
for the seed particles is equal to the energy loss rate for the
background gas at the isothermal shock location.

To close the system, we specify the radial variation of the spatial
diffusion coefficient $\kappa$ by following Le \& Becker (2004, 2005,
2007), who adopted the general form
\begeq
\kappa(r) = \kappa_0 \, u(r) \, \rs \left({r \over \rs} - 1\right)^2
\ ,
\label{eq14}
\fineq
where $\kappa_0$ is a dimensionless positive constant. The value of
$\kappa_0$ is determined self-consistently using the energy conservation
relation $\Qjet = \Qesc$, where the power in the escaping particles,
$\Qesc$, is computed using $\Qesc = \Nesc \, E_*$, with $\Nesc$ denoting
the escape rate for the relativistic particles diffusing out of the disk
at the shock radius $r_*$ with mean energy $E_* = U(r_*) / n(r_*)$. The
spatial diffusion coefficient $\kappa$ is related to the magnetic mean
free path via the usual expression $\kappa = c \, \lambda_{\rm mag}/3$,
where $\lambda_{\rm mag}$ denotes the coherence length of the magnetic
field. Analysis of the three-dimensional random walk of the escaping
particles then yields for the escape parameter $A_0$ the result (Le \&
Becker 2005) $A_0 = (3 \kappa_*)^2/(c \, H_*)^2$, where $\kappa_* \equiv
(\kappa_- + \kappa_+)/2$.

For values of the particle energy $E > E_0$, the source term in
equation~(\ref{eq12}) vanishes and the remaining equation is separable
in energy and space using the functions
\begeq
f_{_n}(E,r) = \left(E \over E_0\right)^{-\lambda_n} \varphi_n(r)
\ ,
\label{eq15}
\fineq
where $\lambda_n$ are the eigenvalues, and the spatial eigenfunctions
$\varphi_n(r)$ satisfy the second-order ordinary differential equation
\begeq
- H u {d \varphi_n \over d r} = {\lambda_n \over 3 r}
{d \over dr} (r H u) \, \varphi_n + {1 \over r} {d \over dr}
\left(r H \kappa \, {d \varphi_n \over d r}\right)
- A_0 \, c \, H_* \, \delta(r-r_*) \, \varphi_n
\ .
\label{eq16}
\fineq
The eigenvalues $\lambda_n$ are determined by applying suitable boundary
conditions to the spatial eigenfunctions, as discussed by Le \&
Becker(2007). The eigenvalues $\lambda_n$ and eigenfunctions
$\varphi_n(r)$ can be determined numerically by computing $H(r)$ and
$\kappa(r)$ using equations~(\ref{eq2}) and (\ref{eq14}), respectively,
once a numerical solution for the inflow speed $u(r)$ has been obtained
(Le \& Becker 2005).

We can establish several useful general properties of the eigenfunctions
by rewriting equation~(\ref{eq16}) in the Sturm-Liouville form
\begeq
{d \over dr}\left[S(r) \, {d\varphi_n \over dr}\right]
+ \lambda_n \, \omega(r) \, \varphi_n(r)
= {A_0 \, c \, S(r) \, \varphi_n(r) \, \delta(r-r_*) \over \kappa(r)}
\ ,
\label{eq17}
\fineq
where
\begeq
\omega(r) \equiv {S \, u \over 3 \kappa} \, {d \ln(r H u)
\over dr}
\ ,
\ \ \ \ \ 
S(r) \equiv {r H \kappa \over r_* H_* \kappa_*}
\ \exp\left[{\rs \over \kappa_0 (r_*-\rs)}-{\rs \over \kappa_0 (r-\rs)}
\right]
\ .
\label{eq18}
\fineq
Using standard manipulations along with the asymptotic behaviors of
$S(r)$ and $\varphi_n(r)$, one can show that the spatial eigenfunctions
form an orthogonal set. It is therefore possible to express the Green's
function $\green$ using an expansion of the form
\begeq
\green(E,r) = \sum_{n=0}^\infty
\, C_n \left(E \over E_0\right)^{-\lambda_n} \varphi_n(r)
\ ,
\label{eq19}
\fineq
where the expansion coefficients $C_n$ are easily computed based on the
orthogonality of the eigenfunctions $\varphi_n(r)$.

\begin{figure}
\begin{center}
\epsfig{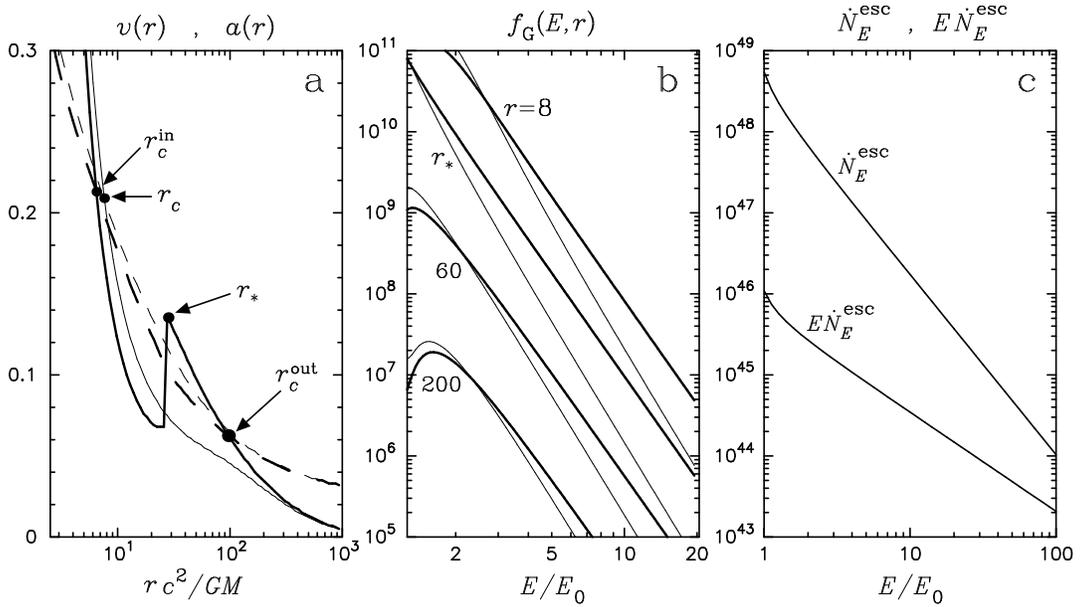}
\end{center}
\caption{Self-consistent results for the disk structure and the particle
transport obtained using the M87 parameters. ({\it a}) Inflow speed
$u(r)$ ({\it solid lines}) and sound speed $a(r)$ multiplied by
$[2\gamma/(\gamma+1)]^{1/2}$ ({\it dashed lines}) plotted as functions
of radius for shocked and smooth (shock-free) disks. ({\it b}) The
Green's function particle distribution (eq.~[\ref{eq19}]) plotted as
functions of energy at various radii inside the disk in cgs units. ({\it
c}) The number and energy distributions for the relativistic particles
escaping from the disk at the shock location (eq.~[\ref{eq20}]) plotted
as a function of energy in cgs units. In panels ({\it a}) and ({\it b}),
the heavy lines denote the shocked disk solutions and the thin lines
represent the shock-free solutions.}
\end{figure}

\section{APPLICATION TO M87}

In our numerical applications to M87, we adopt the observational values
$M \sim 3 \times 10^9 {\rm \ M_{\odot}}$ (e.g., Ford et al. 1994), $\dot
M \sim 1.3 \times 10^{-1} {\rm \ M_{\odot} \ yr^{-1}}$ (Reynolds et al.
1996), and $\Qjet = 5.5 \times 10^{43}\,{\rm erg \ s^{-1}}$ (Reynolds et
al. 1996; Bicknell \& Begelman 1996; Owen, Eilek, \& Kassim 2000). We
utilize natural gravitational units, with $GM=c=1$ and $\rs=2$, except
as noted, and we set $\gamma=1.5$ (see Narayan et. al 1997). In
principle, our model can accommodate any value for the viscosity
parameter $\alpha$. We set $\alpha=0.1$ here in order to demonstrate
that shocks can exist in ADAF disks even in the presence of substantial
viscosity. The remaining parameter values implied by the observations of
M87 are $\epsilon_-=0.001516$, $\epsilon_+=-0.005746$, $j=2.6257$,
$\kappa_0=0.01632$, $\dot N_0=2.75 \times 10^{46} \, {\rm s}^{-1}$,
$\dot N_{\rm esc}=5.81 \times 10^{45} \, {\rm s}^{-1}$, $A_0=0.0153$,
$n(r_*)=9.52 \times 10^{43}\, {\rm cm}^{-3}$, $U(r_*)=8.94 \times
10^{41} \, {\rm erg \ cm}^{-3}$, $E_*/E_0=4.70$, $r_*=26.329$, $r_c^{\rm
in}=6.462$, $r_c^{\rm out}=96.798$, ${\cal M}_-=1.43$, and $H_*=12.10$.
The results obtained for the inflow speed $u$ and the isothermal sound
speed $a$ in a shocked disk are plotted in Figure~1{\it a}. The value of
$\ell$ merges with the Keplerian value $\ellK$ (eq.~[\ref{eq3}]) at
$r=4,658$, which is the outer edge of the ADAF region. Figure~1{\it a}
also includes the dynamical results obtained for $u$ and $a$ in a smooth
(shock-free) disk with $\epsilon_-=\epsilon_+=0.001516$, $j=2.3988$,
$\kappa_0=0.01632$, $\dot N_0=2.75 \times 10^{46} \, {\rm s}^{-1}$,
$\dot N_{\rm esc}=0$, $A_0=0$, $n(r_*)=9.58 \times 10^{43}\, {\rm
cm}^{-3}$, $U(r_*)=2.92 \times 10^{41} \, {\rm erg \ cm}^{-3}$,
$E_*/E_0=1.53$, $r_c=7.572$, and $H_*=13.61$. For the purposes of
comparison with the shocked case, we leave the source located at $r =
26.329$. The value of $j$ for the smooth solution is selected so that
$\ell=\ellK$ at the same radius as in the shocked disk. Our results
represent the first dynamical solutions for ADAF disks with isothermal
shocks and a significant level of viscosity ($\alpha=0.1$). Gu \& Lu
(2001, 2004) obtained solutions for ADAF disks containing
Rankine-Hugoniot shocks, but such shocks have conserved energy transport
rates, and are therefore not relevant for producing outflows. Note that
the mean energy of the relativistic particles at the source radius,
$E_*/E_0$, is significantly larger in the shocked disk than in the
smooth flow.

In Figure~1{\it b} we plot the Green's function energy distribution for
the accelerated particles measured at various radii inside the shocked
and smooth disks. As expected, the presence of the shock results in a
much flatter (i.e. harder) energy spectrum for the accelerated
particles. In Figure~1{\it c}, we plot the number distribution $\dot
N^{\rm esc}_E$ and the energy distribution $E \dot N^{\rm esc}_E$ for
the particles escaping from the shocked disk, computed using (Le \&
Becker 2007)
\begeq
\dot N^{\rm esc}_E(E) = (4 \pi E)^2 \, r_* H_* \, c \, A_0 \green(E,r_*)
\ .
\label{eq20}
\fineq
The total power in the escaping particles, computed using
$\Qesc=\int_{E_0}^\infty E \dot N^{\rm esc}_E\,dE$, is found to equal
$\Qjet = 5.5 \times 10^{43}\,{\rm erg \ s^{-1}}$, which is a useful
self-consistency check on the model.

\section{CONCLUSIONS}

In this Letter, we have obtained for the first time dynamical solutions
for viscous ADAF disks containing shocks based on a relatively large
value for the standard Shakura-Sunyaev viscosity parameter,
$\alpha=0.1$. Utilizing a rigorous mathematical approach, we have
computed the Green's function energy distribution for the relativistic
particles accelerated in the disk, for both shocked and smooth disks.
The results confirm that viscous disks with shocks are much more
efficient particle accelerators than smooth disks. We conclude that the
presence of a shock is an essential ingredient in the formation of the
observed outflows. The absence of strong outflows from luminous X-ray
sources probably reflects the fact that the gas is too dense to allow
efficient Fermi acceleration of a relativistic particle population in
the disk. This helps to explain the observed anticorrelation between
X-ray luminosity and radio/outflow strength (e.g., Reynolds et al.
1996). Our results establish that the luminosity of the M87 jet can be
understood within the context of the disk/shock/outflow model. However,
collimation effects and radiative losses must also be considered in
order to understand the subsequent propagation of the jet through the
extragalactic environment.

\end{document}